\documentclass[aps,prl,showpacs,10pt]{revtex4-1}

\usepackage{dcolumn}
\usepackage{graphicx}
\usepackage{amssymb,amsmath}
\usepackage{natbib}
\usepackage{subfigure}
\usepackage{epstopdf}
\usepackage{placeins}
\usepackage{color}
\usepackage{tikz}
\newcommand{\tpr}{\text{Pr}}
\newcommand{\tra}{\text{Ra}}
\newcommand{\tnu}{\text{Nu}}
\newcommand{\tre}{\text{Re}}

\begin{document}

\title{A pencil distributed finite difference code for strongly turbulent wall-bounded flows}
\author{Erwin P. van der Poel$^{1,}$\footnote{Corresponding author. E-mail address: e.p.vanderpoel@utwente.nl. Phone number: $+31 53 489 2470$}, Rodolfo Ostilla-M\'onico$^{1}$, John Donners$^{2}$ and Roberto Verzicco$^{3,1}$}
\affiliation{
$^1$Department of Physics, Mesa+ Institute, and J.\ M.\ Burgers Centre for Fluid Dynamics, University of Twente, 7500 AE Enschede, The Netherlands \\
$^2$SURFsara, Science Park, 1098 XG Amsterdam, The Netherlands \\
$^3$Dipartimento di Ingegneria Industriale, University of Rome ``Tor Vergata'', Via del Politecnico 1, Roma 00133, Italy}

\date{\today}

\begin{abstract}
We present a numerical scheme geared for high performance computation of wall-bounded turbulent flows. The number of all-to-all communications is decreased to only six instances by using a two-dimensional (pencil) domain decomposition and utilizing the favourable scaling of the CFL time-step constraint as compared to the diffusive time-step constraint. As the CFL condition is more restrictive at high driving, implicit time integration of the viscous terms in the wall-parallel directions is no longer required. This avoids the communication of non-local information to a process for the computation of implicit derivatives in these directions. We explain in detail the numerical scheme used for the integration of the equations, and the underlying parallelization. The code is shown to have very good strong and weak scaling to at least 64K cores.
\end{abstract}

\maketitle

\section{Introduction}

Turbulence is known as the ``last unsolved problem of classical physics''. Direct numerical simulations (DNS) provide a valuable tool for studying in detail the underlying, and currently not fully understood, physical mechanisms behind it.  Turbulence is a dynamic and high dimensional process, in which energy is transferred (cascades) from large vortices into progressively smaller ones, until the scale of the energy is so small that they are dissipated by viscosity. DNS requires solving all of the flow scales, and to adequately simulate a system with very large size separation between the largest and the smallest scale, immense computational power is required. 

The seminal works on homogeneous isotropic turbulence by Orszag \& Patterson \cite{ors72} and on pressure-driven flow between two parallel plates (also known as channel flow) by Kim, Moin and Moser \cite{kim87}, while difficult back then, could be performed easily on contemporary smartphones. Computational resources grow exponentially, and the scale of DNS has also grown, both in memory and floating point operations (FLOPS). In approximately 2005, the clock speed of processors stopped increasing, and the focus shifted to increasing the number of processors used in parallel. This presents new challenges for DNS, and efficient code paralellization is now essential to obtaining (scientific) results.

Efficient paralellization is deeply tied to the underlying numerical scheme. A wide variety of these schemes exist; for trivial geometries, i.e. domains periodic in all dimensions, spectral methods are the most commonly used \cite{kan03}. However, for the recent DNS of wall-bounded flows, a larger variation of schemes is used. For example, in the present year, two channel flow DNSs at similar Reynolds numbers detailed DNSs were performed using both a finite-difference schemes (FDS) in the case of Ref.~\cite{ber14} or a more complex spectral methods in the case of Ref.~\cite{loz14}. FDS also present several advantages, they are very flexible, allowing for complex boundary conditions and/or structures interacting through the immersed boundary method with relative ease \cite{fad00}. A commonly asserted disadvantage of low-order FDS is the higher truncation error relative to higher order schemes and spectral methods. However, this is only true in the asymptotic limit where the grid spacing $\Delta x\to0$ that is commonly not reached. Additionally, aliasing errors are much smaller for lower order schemes. Lower-order schemes have been shown to produce adequate first- and second-order statistics, but require higher resolution when compared to spectral methods for high order statistics \cite{rai91,dup08,vre14}. Because lower-order schemes are computationally very cheap the grid resolution can in general be larger for the same computational cost compared to higher order schemes, although one has to consider the higher memory bandwidth over FLOPS ratio. 

In this manuscript, we will detail the parallelization of a second-order FDS based on Verzicco \& Orlandi \cite{ver96} to two wall bounded systems, Rayleigh-B\'{e}nard (RB) convection, the flow in a fluid layer between two parallel plates; one heated from below and cooled from above and Taylor-Couette (TC) flow, the flow between two coaxial independently rotating cylinders, although our code can easily extended to any flow that is wall-bounded in one dimension. This FDS scheme has already been used in pure Navier-Stokes simulations \cite{ver96}, with immersed boundary methods \cite{str06}, for Rayleigh-B\'enard convection \cite{ver03,ver08,kun08,ste11,poe13,poe14,ste14} and for Taylor-Couette flow \cite{ost13,ost14}. The numerical results have been validated against experimental data numerous times.  We will exploit several advantages of the large $\tre$ regime and the boundary conditions to heavily reduce communication cost;  opening the possibility to achieve much higher driving.

The manuscript is organized as follows: Section 2 describes TC and RB in more detail. Section 3 details the numerical scheme used to advance the equations in time. Section 4 shows that in thermal convection, the Courant-Friedrichs-Lewy (CFL) \cite{cou28} stability constraints on the timestep due to the viscous terms become less strict than those due to the non-linear terms at high Rayleigh (Reynolds) numbers. Section 5 details a pencil decomposition to take advantage of the new time integration scheme and the choice of data arrangement in the pencil decomposition. Finally, section 6 compares the computational cost of existing and the new approach and presents an outlook of what further work can be done to combine this approach with other techniques. 

\section{Rayleigh-B\'{e}nard convection and Taylor-Couette flow}

\begin{figure}
    \centering
	\raisebox{.2\height}{\includegraphics[trim=0cm 0cm 0cm 0cm,width=0.40\textwidth]{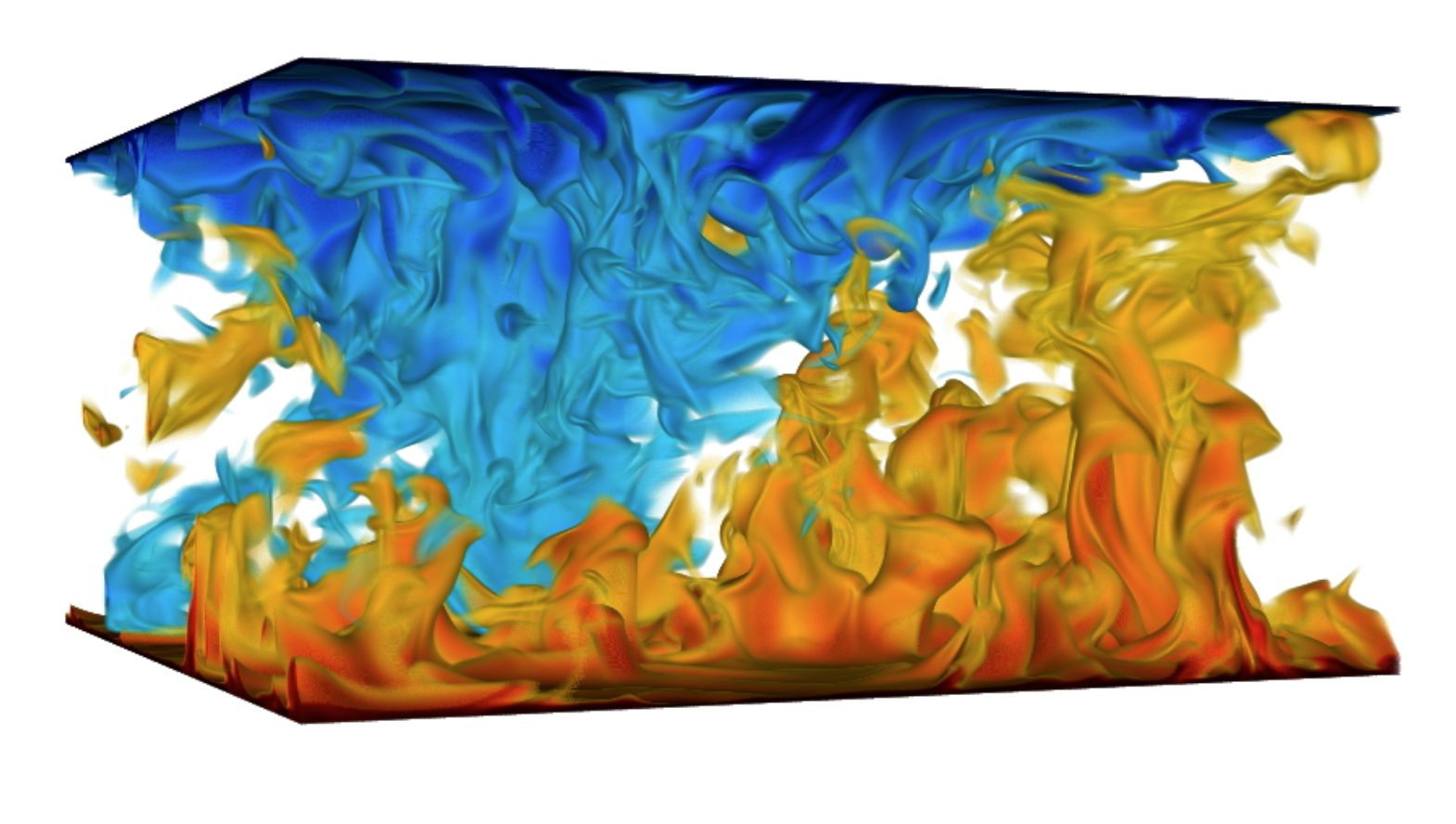}}
	\hspace{30 mm}
 	\includegraphics[trim=0cm 0cm 0cm 0cm,width=0.20\textwidth]{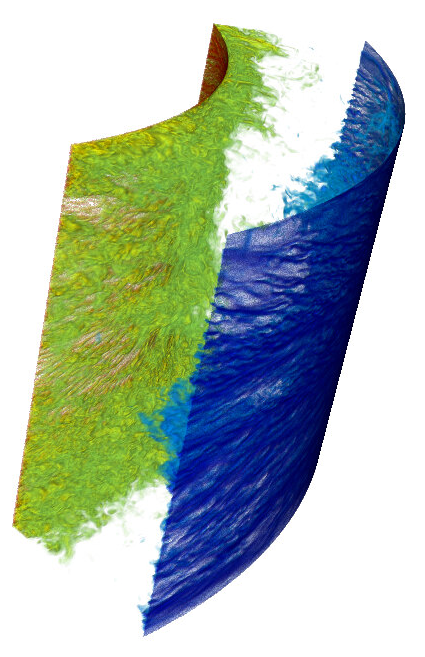}
    \caption{ Left: RB flow for $\tra=10^8$, $\tpr=1$ and $\Gamma=2$ in Cartesian coordinates. The horizontal directions are periodic and the plates are subjected to a no-slip and isothermal boundary condition. Red/yellow indicates hot fluid, while (light)blue indicates cold fluid. The small heat carrying structures known as thermal plumes as well as a large scale circulation can be seen in the visualization, highlighting the scale separation in the flow. Right: TC flow with an inner cylinder Reynolds number $\tre=10^5$, a stationary outer cylinder, and a radius ratio $\eta=r_i/r_o=0.714$. Green fluid has a high angular velocity while blue fluid has a low angular velocity. The smallness of the structures responsible for torque transport, and thus the need for fine meshes, can be appreciated clearly. }
\label{fig:snapshots}
\end{figure}

RB and TC are paradigmatic models for convective and shear flows, respectively. They are very popular systems because they are mathematically well defined, experimentally accessible and reproduce many of the interesting phenomena observed in applications. A volume rendering of the systems can be seen in Fig. \ref{fig:snapshots}. The Reynolds numbers in the common astro- and geo-physical applications are much higher than what can be reached currently in a laboratory. Therefore it is necessary to extrapolate available experimental results to the large driving present in stars and galaxies. This extrapolation becomes meaningless when transitions in scaling behaviour are present, and it is expected that once the Rayleigh number, i.e. the non-dimensional temperature difference, becomes large enough, the boundary layers transition to turbulence. This transition would most likely affect the scaling of interesting quantities. However, experiments disagree on exactly where this transition takes place \cite{roc10,he12}. DNS can be used to understand the discrepancies among experiments. However, to reach the high Rayleigh numbers ($\tra$) of experiments new strategies are required. DNS must resolve all scales in the flow, and the scale separation between the smallest scale and the largest scale grows with Reynolds number. This means larger grids are needed, and the amount of computational work $W$ scales approximately as $W\sim \tre^{4}$ \cite{pop00}. 

Simulations of RB commonly imitate the cylindrical geometry most used in experiments. Recently, a DNS with aspect ratio $\Gamma=D/L=1/3$, where $D$ is the diameter of the plates and $L$ the height of the cell reached $\tra=10^{12}$ using 1.6 Billion points with a total cost of 2 Million CPU hours \cite{poe14b}. DNS in other geometries have been proposed, such as homogeneous RB, where the flow is fully periodic and a background temperature gradient is imposed. This geometry is easy to simulate \cite{loh03}, but presents exponentially growing solutions and does not have a boundary layer, thus not showing any transition \cite{cal06}. Axially homogeneous RB, where the two plates of the cylinder are removed, and the sidewalls kept and a background temperature gradient is imposed to drive the flow has also been simulated \cite{sch12}. This system does not have boundary layers on the plates and does not show the transition. Therefore, it seems necessary to keep both horizontal plates, having at least one wall-bounded direction. The simplest geometry is a parallelepiped box, periodic in both wall-parallel directions, which we will call ``rectangular'' RB for simplicity. Rectangular RB is receiving more attention recently \cite{cal02,par04,sch08b,gay13}, due to possibility to reach higher Ra as compared to more complex geometries. It is additionally the geometry that is closest to natural applications, where there are commonly no sidewalls.

For TC, we have one naturally periodic dimension, the azimuthal extent. The axial extent can be chosen to be either bounded by end-plates, like in experiments, or to be periodic. Axial end-plates have been shown to cause undesired transitions to turbulence if TC is in the linearly stable regime \cite{avi12}, or to not considerably affect the flow if TC is in the unstable regime \cite{gil12}. Large $\tre$ DNS of TC focus on axially periodic TC, bounding the flow only in the radial direction \cite{bra13,ost14}. Therefore, the choice of having a single wall-bounded direction for DNS of both TC and RB seems justified.

\section{Numerical scheme}

The code solves the Navier-Stokes equations with an additional equation for temperature in three-dimensional coordinates, either Cartesian or cylindrical. For brevity, we will focus on the RB Cartesian problem, although all concepts can be directly translated to TC in cylindrical coordinates system by substituting the vertical direction for the radial direction, and the two horizontal directions by the axial and azimuthal directions.

The non-dimensional Navier-Stokes equations with the Boussinesq approximation for RB read:

\begin{equation}
\nabla \cdot \textbf{u} = 0,
\label{eq:Ns1}
\end{equation}

\begin{equation}
\displaystyle\frac{\partial\textbf{u}}{\partial t} + \textbf{u} \cdot \nabla \textbf{u} = -\nabla p + \sqrt\frac{\tpr}{\tra}\nabla^2\textbf{u} + \theta \textbf{e}_x,
\label{eq:Ns2}
\end{equation}

\begin{equation}
\displaystyle\frac{\partial\theta}{\partial t} + \textbf{u}\cdot \nabla \theta =  \sqrt\frac{1}{\tpr\tra}\nabla^2\theta,
\label{eq:Ns3}
\end{equation}

\noindent where $\textbf{u}$ is the non-dimensional velocity, $p$ the non-dimensional pressure, $\theta$ the non-dimensional temperature and $t$ the non-dimensional time. For non-dimensionalization, the temperature scale is the temperature difference between both plates $\Delta$, the length scale is the height of the cell $L$ and the velocity scale is the free-fall velocity $U_f=\sqrt{g\beta\Delta L}$, where $g$ is gravity and $\beta$ is the isobaric expansion coefficient of the fluid. $\tpr=\nu/\kappa$ is the fluid Prandtl number, where  $\nu$ is the kinematic viscosity and $\kappa$ is the thermal conductivity. The Rayleigh number is defined in this case as $\tra=g\beta\Delta L^3/\nu\kappa$. Finally, $\textbf{e}_x$ is the unitary vector in parallel direction to gravity, which is also the plate-normal direction. 

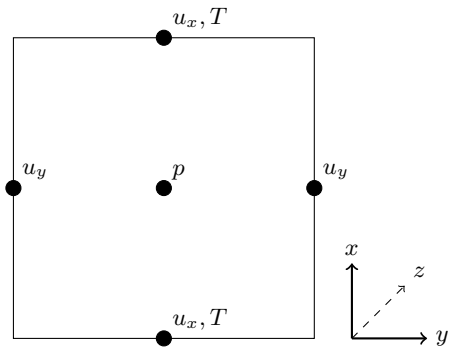
\begin{figure}
\centering
\hspace{1cm}
\begin{tikzpicture}
\draw (0,0) rectangle (4,4);
\filldraw[black] (2,2) circle(1mm);
\node [above right,black] at (2,2) {$p$};
\filldraw[black] (0,2) circle(1mm);
\node [above right,black] at (0,2) {$u_y$};
\filldraw[black] (4,2) circle(1mm);
\node [above right,black] at (4,2) {$u_y$};
\filldraw[black] (2,0) circle(1mm);
\node [above right,black] at (2,0) {$u_x,T$};
\filldraw[black] (2,4) circle(1mm);
\node [above right,black] at (2,4) {$u_x,T$};
\draw [->,thick] (4.5,0) -- (5.5,0);
\draw [->,thick] (4.5,0) -- (4.5,1.0);
\draw [->,thin, dashed] (4.5,0) -- (5.2,0.7);
\node [above,black] at (4.5,1.0) {$x$};
\node [right,black] at (5.5,0) {$y$};
\node [above right,black] at (5.2,0.7) {$z$};
\end{tikzpicture}
\caption{Location of pressure, temperature and velocities of a 2D simulation cell. The third dimension ($z$) is omitted for clarity. As on an ordinary staggered scheme, The velocity vectors are placed on the borders of the cell and pressure is placed in the cell center. The temperature is placed on the same nodes as the vertical velocity, to ensure exact energy conservation.}
\label{fig:disc}
\end{figure}

As mentioned previously, the two horizontal directions are periodic and the vertical direction is wall-bounded. The spatial discretization used is a conservative second-order centered finite difference with velocities on a staggered grid. The pressure is calculated at the center of the cell while the temperature field is located on the $u_x$ grid. This is to avoid the interpolation error when calculating the term $\theta \textbf{e}_x$ in equation \ref{eq:Ns2}. The scheme is energy conserving in the limit $\Delta t\to0$. A two-dimensional (for clarity) schematic of the discretization is shown in Fig.~\ref{fig:disc}. For the case of thermal convection an additional advantage of using FDS is present: the absence of pressure in the advection/diffusion equation for scalars causes shock-like behaviour to appear in the temperature (scalar) field \cite{ost14d} and low-order schemes fare better in this situation.

Time marching is performed with a fractional-step third-order Runge-Kutta (RK3) scheme, in combination with a Crank-Nicholson scheme \cite{rai91} for the implicit terms. A second-order Adams-Bashforth (AB2) method is additionally implemented. However, in all production runs the RK3 method takes precedence over AB2 even though the total RK3 time step includes three substeps as compared to one for AB2. The theoretical stability limit of AB2 and RK3 are CFL numbers lower than $1$ and $\sqrt{3}$, respectively. In practice, the maximum CFL numbers of AB2 and RK3 are approximately $0.3$ and $1.3$, respectively. Because of three times higher amount of substeps in RK3, the computational cost is proportionally higher compared to AB2. Nevertheless, RK3 is more efficient as the progression in physical time per computional cost is better. In addition, even though the Crank-Nicholson integration with $\mathcal{O}([\Delta t]^2)$ error is the weakest link, the $\mathcal{O}([\Delta t]^3)$ error of RK3 decreases the total error significantly compared to the $\mathcal{O}([\Delta t]^2)$ error of AB2. In addition, RK3 is self--starting at each time step without decreasing the accuracy and without needing additional information in the restart file. AB2 would require two continuation files per quantity. 

The pressure gradient is introduced through the ``delta'' form of the pressure \cite{lee01}: an intermediate, non-solenoidal velocity field $\textbf{u}^*$ is calculated using the non-linear, the viscous and the buoyancy terms in the Navier-Stokes equation, as well as the pressure at the current time sub-step: 

\begin{equation}
\displaystyle\frac{\textbf{u}^*-\textbf{u}^j}{\Delta t} = \left [ \gamma_l H^j + \rho_l H^{j-1} - \alpha_l \mathcal{G} p^j + \alpha_l(\mathcal{A}^j_x + \mathcal{A}^j_y + \mathcal{A}^j_z) \displaystyle\frac{(\textbf{u}^*+\textbf{u}^j)}{2} \right ],
\end{equation}

\noindent where the superscript $j$ denotes the substep, $\mathcal{A}_i$ is the discrete differential relationship for the viscous terms in the $i$th-direction, $\mathcal{G}$ the discrete gradient operator and $H^j$ all explicit terms. The coefficients $\gamma_l$, $\rho_l$ and $\alpha_l$ depend on the time marching method used. The pressure required to enforce the continuity equation at every cell is then calculated by solving a Poisson equation for the pressure correction $\phi$:

\begin{equation}
 \nabla^2 \phi = \displaystyle\frac{1}{\alpha_l \Delta t} (\nabla \cdot \textbf{u}^*),
 \label{eq:pc}
\end{equation}

\noindent or in discrete form:

\begin{equation}
  \mathcal{L}\phi = \displaystyle\frac{1}{\alpha_l \Delta t} (\mathcal{D}\textbf{u}^*),
\end{equation}

\noindent where  $\mathcal{D}$ the discrete divergence operator, and $\mathcal{L}$ is the discrete Laplacian operator, $\mathcal{L}=\mathcal{D}\mathcal{G}$. The velocity and pressure fields are then updated using: 

\begin{equation}
 \textbf{u}^{j+1} = \textbf{u}^* - \alpha_l \Delta t ( \mathcal{G} \phi),
 \label{eq:correctedu}
\end{equation}

\noindent and 

\begin{equation}
 p^{j+1} = p^j + \phi - \frac{\alpha_l \Delta t}{2\tre} (\mathcal{L} \phi),
  \label{eq:correctedp}
\end{equation}

\noindent making $\textbf{u}^{j+1}$ divergence free. 

\begin{figure}
    \centering
    \includegraphics[width=0.45\textwidth]{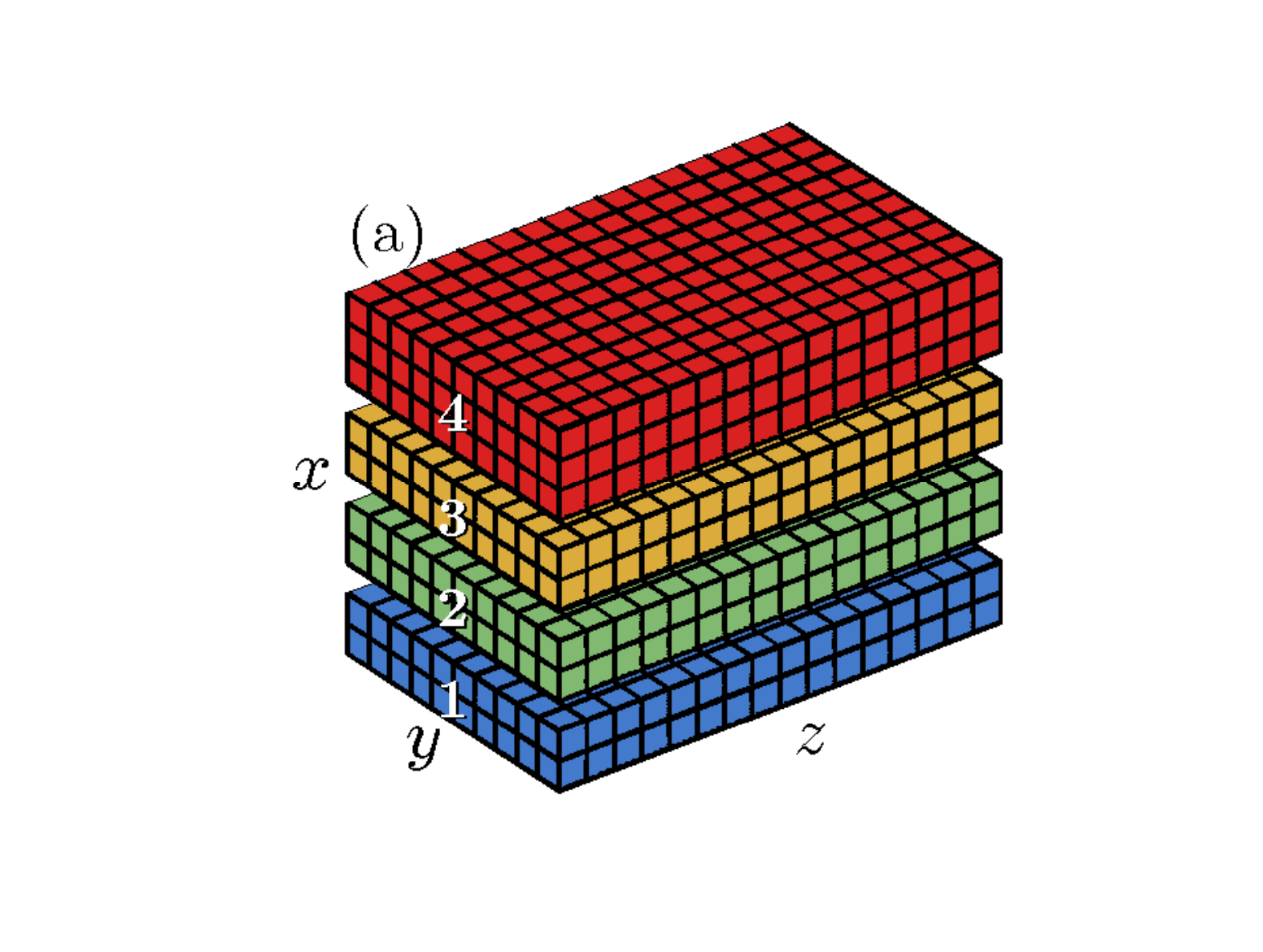}
    \includegraphics[width=0.45\textwidth]{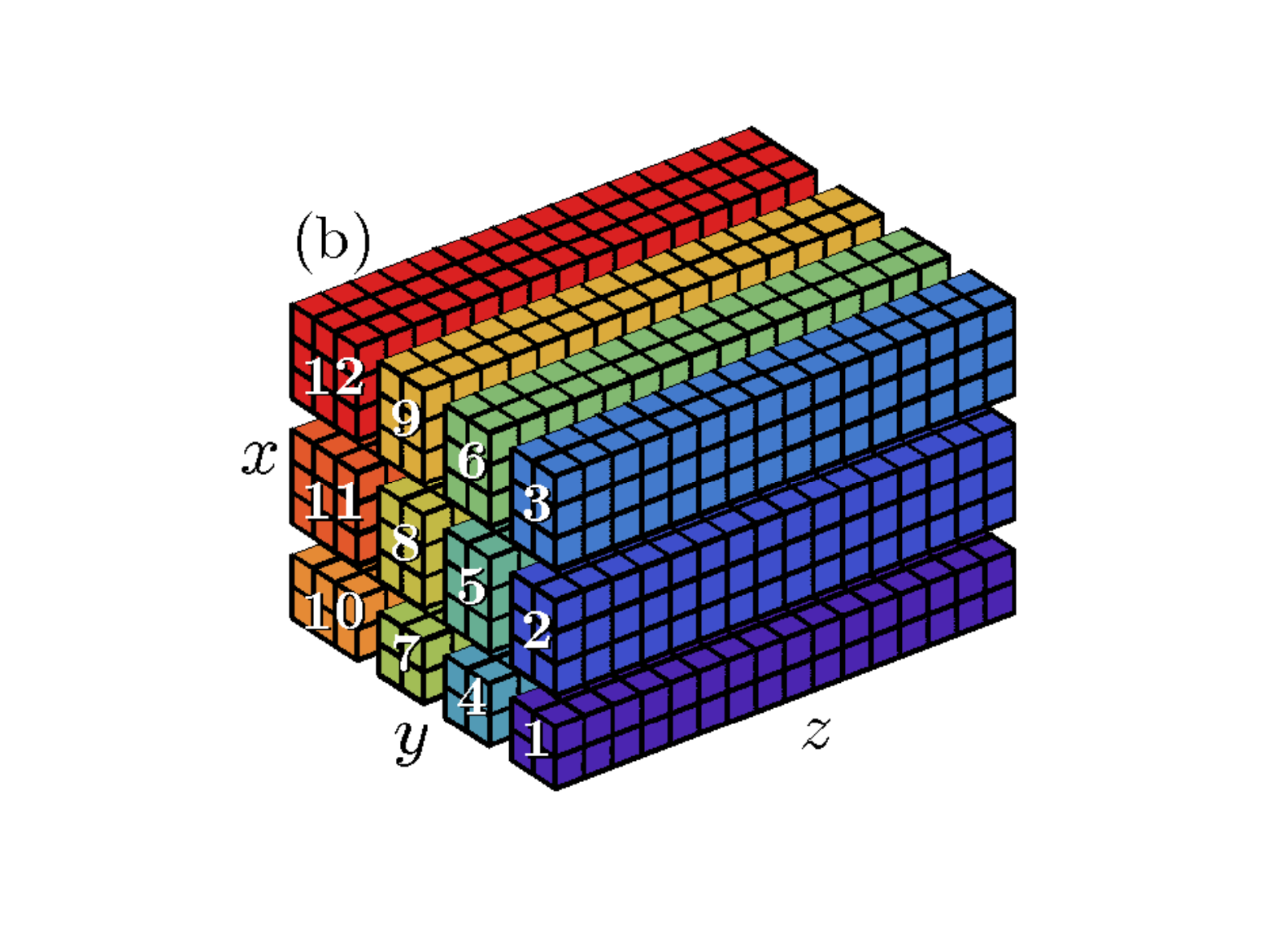}
    \caption{Left panel: Slab-type domain decomposition using four MPI processes. Right panel: Pencil-type domain decomposition using twelve MPI processes. }
\label{fig:mpidecomp}
\end{figure}

The original numerical scheme treats all viscous terms implicitly. This would result in the solution of a large sparse matrix, but this is avoided by an approximate factorization of the sparse matrix into three tridiagonal matrices; one for each direction \cite{ver96}. The tridiagonal matrices are then solved using Thomas' algorithm, (with a Sherman-Morrison perturbation if the dimension is periodic), in $\mathcal{O}(\text{N})$ time. The calculation is thus simplified at the expense of introducing an error $\mathcal{O}(\Delta t^3)$. This method was originally developed and used for small Reynolds number problems, and without having in mind that data communication between different processes could be a bottleneck. The first parallelization scheme with MPI was a 1D-domain ``slab'' decomposition, visualized in the left panel of Fig.~\ref{fig:mpidecomp}. The main bottlenecks were found in the all-to-all communications present in the pressure-correction step and the tridiagonal solver in the direction in which the domain is decomposed (cf. Table \ref{tb:comms} for more details). Slab decompositions are easy to implement, but are limited in two ways: First, the number of MPI processes cannot be larger than N, the amount of grid points in one dimension. A hybrid MPI-OpenMP decomposition can take this limit further, but scaling usually does not go further than $10^4$ cores. Second, the size of the ``halo'' (or ghost) cells becomes very significant with increasing number of cores. Halo cells are cells which overlap the neighbour's domain, and whose values are needed to compute derivatives. In the limit of one grid point per processor, halo cells are of the size of the domain in a second-order scheme, and even larger for higher order schemes. 

2D-domain decompositions, also known as ``pencil'' decompositions mitigate these problems. A schematic of this decomposition is shown in Fig.~\ref{fig:mpidecomp}. To implement this decomposition, the 2DECOMP \cite{li10} library has been used, and extended upon to suit the specifics of our scheme. The limit on the amount of processes is now raised to $\text{N}^2$, and the size of the halo cells on every core \emph{decreases} with increasing amount of cores, so the amount of communication per core decreases. However, for a pencil decomposition solving all the tridiagonal matrices requires all-to-all communications for two directions, instead of one direction in the case of slab decomposition. As mentioned previously, solving the tridiagonal matrices involves large data communication, especially in the context of pencil decompositions. In this manuscript we attempt to eliminate them as far as possible and at the same time implement an efficient pencil domain decomposition by arranging the data to gain advantage of the inherent anisotropy with respect to the grid point distributions.

\section{Constraints on the timestep}

In the new scheme, solving the tridiagonal matrices in the horizontal directions is avoided by integrating not only the advection terms but also the viscous terms explicitly. A major concern is that this can cause the temporal stability issues that the implicit integration used to negate. In this section we argue that for high Ra, using the Courant-Frederich-Lewy (CFL) \cite{cou28} time step size constraint is sufficient. The CFL condition ensures the stability of the integration of the advection terms, and for high Ra it additionally ensures stability of the viscous terms. The grid point distribution in the wall-normal direction is different compared to the periodic direction. Namely, it is non-uniform in the wall-normal direction, with clustering of points near the boundaries in order to adequately resolve the boundary layers. As the periodic directions are homogeneous, no such clustering is required and the grid point distribution can be uniform. As a consequence, the minimum grid spacing in the wall-normal direction is much smaller than in the horizontal directions. Because of the scaling of both the time-step constraints, the viscous terms in the wall-normal direction do require implicit integration for all Ra.

In this semi-implicit method, both viscous terms and the non-linear terms are integrated explicitly. This requires two stability constraints on the time-step: one due to the non-linear terms, and one due to the horizontal viscous terms. For the non-linear terms, the definition of the CFL condition is given by:

\begin{equation}
\Delta t_{\textbf{u}\cdot\nabla\textbf{u}} \leq \mathcal{C}_1 \min_{\forall \textbf{x} \in X} \frac{1}{\frac{|u_x|(\textbf{x})}{\Delta x(\textbf{x})}+\frac{|u_y|(\textbf{x})}{\Delta y(\textbf{x})}+\frac{|u_z|(\textbf{x})}{\Delta z(\textbf{x})}},
\label{eq:cfl}
\end{equation}
where $\mathcal{C}_1$ is the integration scheme dependent Courant number, $\textbf{x}$ is the position vector, $X$ is the complete domain and $|\cdot|$ gives the absolute value. Here $\Delta x$ gives the (non-dimensional) grid spacing in the $x$ direction at position $\textbf{x}$. The wall-normal direction is $x$ and the wall parallel directions are $y$ and $z$ (cf. Fig.~\ref{fig:disc}). 

The additional constraint originates from the viscous terms, and is given by:

\begin{equation}
\Delta t_{\nu\nabla^2\textbf{u}} \leq \tre~\mathcal{C}_2 \min_{\forall \textbf{x} \in X}  (\Delta y(\textbf{x})+\Delta z(\textbf{x}))^2,
\label{eq:viscouscfl}
\end{equation}
where $\tre = \sqrt{\tra/\tpr}$ is the Reynolds number, and $\mathcal{C}_2$ a number which depends on the integration scheme and the number of dimensions treated explicitly. This condition only needs to be satisfied in the horizontal directions, and not in the vertical direction, as the time integration of the vertical second derivatives is kept implicit. Thus $\Delta x(\textbf{x})^2$ can be very small, and the resulting time-step would make the simulation infeasible. 

We can now compare the two CFL constraints, and show that the non-linear constraint is more restrictive than the viscous constraint in the homogeneous directions. As $|u_z|$ and $\Delta z$ are strictly positive, we have:

\begin{equation}
\frac{\tre~\mathcal{C}_2}{\frac{|u_x|(\textbf{x})}{\Delta x(\textbf{x})}+\frac{|u_y|(\textbf{x})}{\Delta y(\textbf{x})}+\frac{|u_z|(\textbf{x})}{\Delta z(\textbf{x})}} < \frac{\tre~\mathcal{C}_2}{\frac{|u_y|(\textbf{x})}{\Delta y(\textbf{x})}+\frac{|u_z|(\textbf{x})}{\Delta z(\textbf{x})}}.
\label{eq:spacingverthort}
\end{equation}
Including the wall-normal grid spacing in the CFL condition gives a smaller time-step than only using horizontal spacing, and thus the expression on the right is an upper bound on the time-step. If we then use that the grid is uniformly spaced, and equally spaced in both horizontal directions, we can simplify the expression as $\Delta y(\textbf{x})= \Delta z(\textbf{x}) = \Delta y$.  We also know that the dimensionless velocity is $|u| \sim \mathcal{O}(1)$ by normalization. Using all of this, we obtain:

\begin{equation}
\Delta t_{\textbf{u}\cdot\nabla\textbf{u}}  \sim \mathcal{O}\big(\Delta y\big),
\label{eq:cflscale}
\end{equation}
for the non-linear CFL condition and

\begin{equation}
 \Delta t_{\nu\nabla^2\textbf{u}} \sim \mathcal{O}\big(\tre~\Delta y^2\big),
\label{eq:viscouscflscale}
\end{equation}

\noindent from the CFL criterion for the viscous terms. If we assume $Pr \sim \mathcal{O}(1)$, we can get a bound on the viscous time-step as a function of $\tra$, $\Delta t_{\nu\nabla^2\textbf{u}} \sim \mathcal{O}\big(\tra^{\frac{1}{2}}\Delta y^2\big)$.

To compare both bounds, we need an estimation for $\Delta y$. For a resolved DNS, $\Delta y$ should be similar to the smallest physical length scale in the system. Several length scales can be chosen in the thermal convection problem. The first choice stems from homogeneous turbulence, where the most commonly used length scale that determines the numerical resolution is the Kolmogorov length scale, $\eta_K = \nu^{3/4}\epsilon^{-1/4}$, where $\epsilon$ is the viscous dissipation rate. For RB, we can obtain an estimate for the Kolmogorov scale by using that the volume and time averaged viscous dissipation rate can be expressed directly as a function of $\tnu$, $\tra$ and $\tpr$ \cite{shr90,sig94}:

\begin{equation}
\langle \epsilon \rangle_{V,t} = \frac{\nu^3}{L^4}\tra\tpr^{-2}(\tnu-1).
\label{eq:viscdiss}
\end{equation}

For high Ra simulations, $\tnu \gg 1$. Using again that $\tpr \sim\mathcal{O}(1)$, we can obtain an estimate for $\eta_K$, and thus the grid spacing as $\Delta y = \eta_K/L \approx 1/(\tra\tnu)$. If we assume $\tnu \sim \tra^\gamma$ we can now compare both CFL constraints on the time step, obtaining  $\gamma \leq 1$ as a requirement for the non-linear CFL to be more restrictive on the time step than the viscous CFL constraint. 

In RB convection, another restrictive length scale naturally arises, i.e. that of thermal plumes. These are conceptualized as detaching pieces of thermal boundary layers. The thickness of a thermal boundary layer can be approximated by $\lambda \approx 1/(2\tnu)$. Using this as an estimate for $\Delta y$ in Eq. \ref{eq:cfl}  and Eq. \ref{eq:viscouscfl} gives another bound: $\gamma \leq 1/2$. Trivial upper bounds in RB convection give a physical upper bound of $\tnu \sim \tra^{1/2}$ \cite{spi71}, indicating that for the mild assumptions made, the criteria $\gamma \leq 1/2$ is always satisfied. This signifies that the scaling of $\Delta t_{\textbf{u}\cdot\nabla\textbf{u}}$ is more restrictive than $\Delta t_{\nu\nabla^2\textbf{u}}$, which results in that using only the non-linear CFL constraint in the time-marching algorithm, inherently satisfies the stability constraints imposed by the explicit integration of the horizontal components of the viscous terms. Including the vertical non-uniform grid in this derivation makes this statement even more valid, as the used CFL time step is based on this grid (Eq. \ref{eq:cfl}). Inherent to the big-O-notation is the absorption of the coefficients and offsets. This makes this derivation only valid for high $\tra$ flows. For low $\tra$, the solver will be unstable the viscous constraint is not satisfied in this regime. 

In addition, we note that the previous analysis can be applied for the scalar (temperature) equation as long as $\tpr\sim \mathcal{O}(1)$. If $\tpr \gg 1$, which is the case in some applications, the CFL constraint on the horizontal \emph{conductive} terms becomes $\Delta t_{\kappa\nabla^2T} \sim \mathcal{O}\big(\tpr^{\frac{1}{2}}\tra^{\frac{1}{2}} (\Delta y)^2\big)$, which means a stricter restriction on the time-step than Eq. \ref{eq:viscouscflscale}. This means that the $\tra$ of the flow required to make $\Delta t_{\textbf{u}\cdot\nabla\textbf{u}} \leq \Delta t_{\kappa\nabla^2T}$ will be higher.

\section{Code Parallelization}

\begin{figure}
    \centering
    \includegraphics[trim=2cm 2cm 2cm 0cm,clip=true,width=0.325\textwidth]{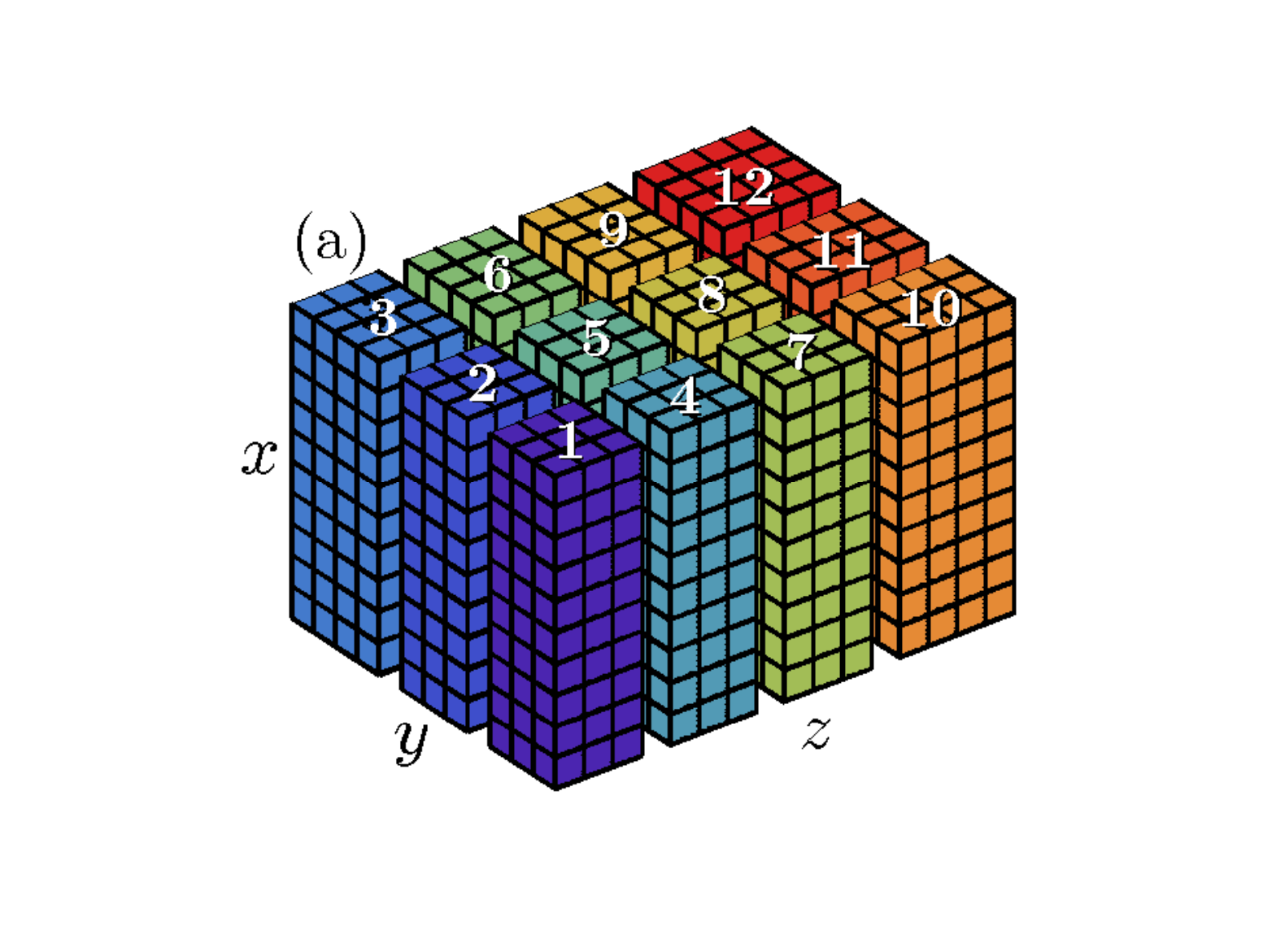}
    \includegraphics[trim=2cm 2cm 2cm 0cm,clip=true,width=0.325\textwidth]{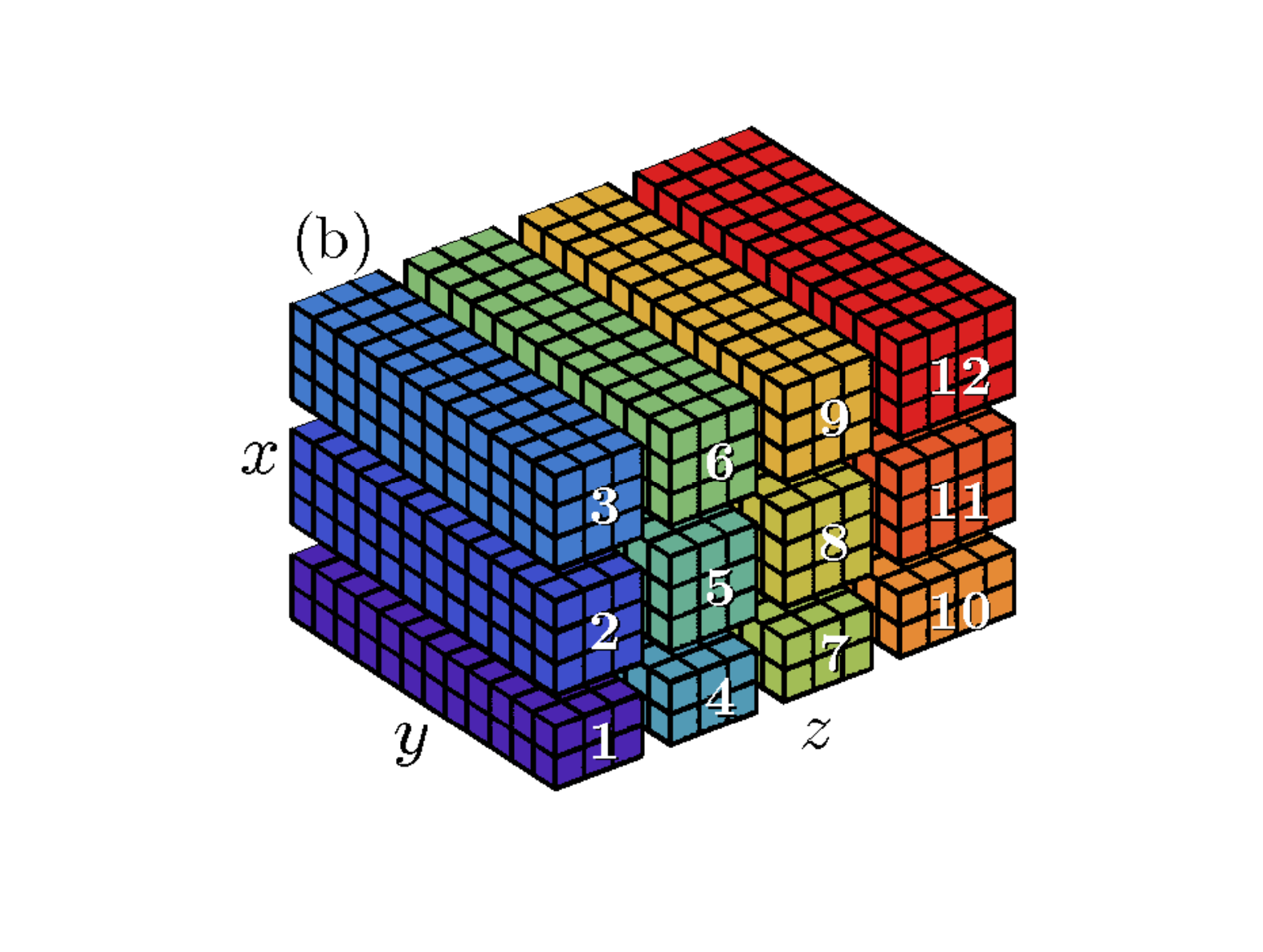}
    \includegraphics[trim=2cm 2cm 2cm 0cm,clip=true,width=0.325\textwidth]{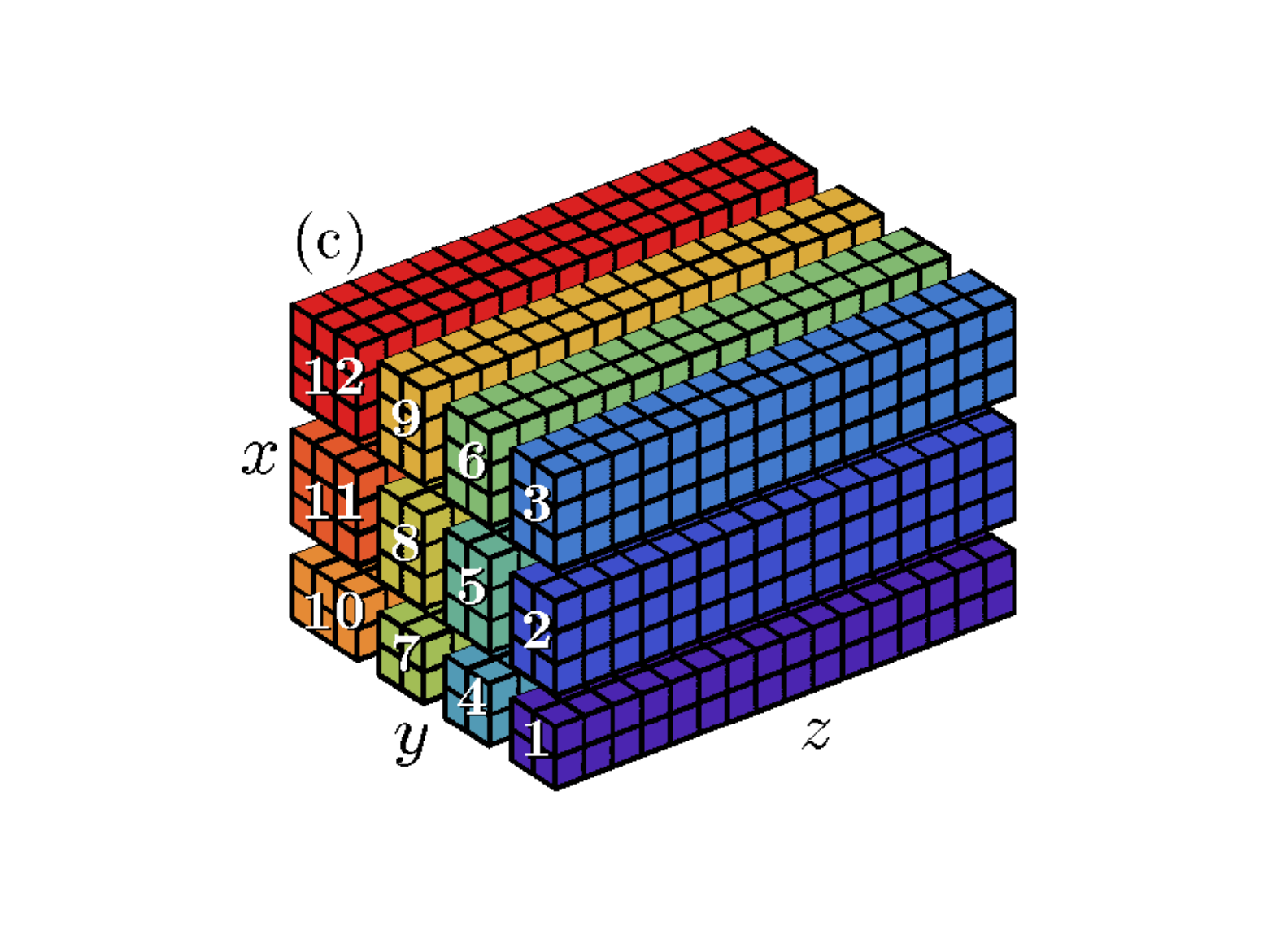}
    \caption{Domain decomposition of a $16 \times 12 \times 10$ grid using $12$ distributed memory processes on a $4 \times 3$ process grid. Only data that is exclusive to one process is shown; i.e. a 1 gridpoint-sized halo is transparent in this figure. The pencils are a) $x$, b) $y$ or c) $z$ oriented.}
\label{fig:trans}
\end{figure}

In the previous section, we reasoned that for large $\tra$ the implicit integration of the viscous terms in the horizontal direction becomes unnecessary. The calculation becomes local in space as the two horizontal directions no longer require implicit solvers to calculate the intermediate velocity field $\textbf{u}^*$. In this case it is worth decomposing the domain such that the pencils are aligned in the wall-normal ($x$) direction, i.e. that every processor possesses data from $x_1$ to $x_N$ (cf. Fig. \ref{fig:trans}). Halo updates must still be performed during the computation of $\textbf{u}^*$, but this memory distribution completely eliminates all the all-to-all communications associated to the viscous implicit solvers, as for every pair ($y$,$z$), a single processor has the full $x$ information, and is able to solve the implicit equation in $x$ for the pair ($y$,$z$) without further communication.

All-to-all communications are unavoidable during the pressure correction step, as a Poisson equation must be solved. As the two wall-parallel directions are homogeneous and periodic, it is natural to solve the Poisson equation using a Fourier decomposition in two dimensions. Fourier transforming variables $\phi$ and the right side in Eq. \ref{eq:pc} reduces the pressure correction equation to:

\begin{equation}
 \left ( \displaystyle\frac{\partial^2}{\partial x^2} - \omega_{y,j}^2 - \omega_{z,k}^2 \right ) \mathcal{F}(\phi) = \mathcal{F}\left[\displaystyle\frac{1}{\alpha_l \Delta t} (\mathcal{D}\textbf{u}^*)\right ]
 \label{eq:pcdisc}
\end{equation}


\noindent where $\mathcal{F}(\cdot)$ denotes the 2D Fourier transform operator, and $\omega_{y,j}$ and $\omega_{z,k}$ denote the $j$-th and $k$-th modified wavenumbers in $y$ and $z$ direction respectively, defined as:

$$
 \omega_{y,j} = \begin{cases}
    \left(1-\cos\left[\displaystyle\frac{2\pi(j-1)}{N_y}\right]\right)\Delta_y^{-2} & : \mbox{for } j \leq \frac{1}{2}N_y + 1 \\
    \left(1-\cos\left[\displaystyle\frac{2\pi(N_y-j+1))}{N_y}\right]\right)\Delta_y^{-2} & : \mbox{otherwise}
\end{cases}
$$

\noindent and $\omega_{z,k}$ is defined in an analogous way. A modified wavenumber is used, instead of the real wavenumber, to prevent that the Laplacian has higher accuracy in some dimensions. In the limit $\Delta y\to 0$, the modified wavenumbers converge to the real wavenumbers.

By using a second order approximation for $\partial^2_x$, the left hand side of the equation is reduced to a tridiagonal matrix, and thus the Poisson equation is reduced to a 2D FFT followed by a tridiagonal (Thomas) solver. This allows for the exact solution of the Poisson equation in a single iteration with $\mathcal{O}(N_x N_y N_z \log[N_y] \log[N_z])$ time complexity. Due to the domain decomposition, several data transposes must be performed during the computation of the equation. The algorithm for solving the Poisson equation is as follows: 

\begin{enumerate}
 \item Calculate $(\mathcal{D}\textbf{u}^*)/(\alpha_l\Delta t)$ from the $x$-decomposed velocities.
 \item Transpose the result of (1) from a $x$-decomposition to a $y$-decomposition.
 \item Perform a real-to-complex Fourier transform on (2) in the $y$ direction.
 \item Transpose (3) from a $y$-decomposition to $z$-decomposition.
 \item Perform a complex-to-complex Fourier transform on (4) in the $z$ direction.
 \item Transpose (5) from a $z$-decomposition to a $x$-decomposition.
 \item Solve the linear system of Eq. \ref{eq:pcdisc} with a tridiagonal solver in the $x$-direction.
 \item Transpose the result of (7) from a $x$-decomposition to a $z$-decomposition.
 \item Perform a complex-to-complex inverse Fourier transform on (8) in $z$ direction.
 \item Transpose (9) from a $z$-decomposition to a $y$-decomposition.
 \item Perform a complex-to-real inverse Fourier transform on (10) in a $y$ direction.
 \item Transpose (11) from a $y$-decomposition to a $x$-decomposition.
\end{enumerate}

The last step outputs $\phi$ in real space, decomposed in $x$-oriented pencils, ready for applying in Eqs \ref{eq:correctedu}-\ref{eq:correctedp}. Once the Poisson equation is solved, the corrected velocities and pressures are computed using Eqs. \ref{eq:correctedu}-\ref{eq:correctedp}. The temperature and other scalars are advected and the time sub-step is completed. The algorithm outlined above only transposes one 3D array, instead of three velocity fields, making it very efficient. Figure \ref{fig:trans} shows a schematic of the data arrangement and the transposes needed to implement the algorithm. We wish to highlight that this algorithm also uses all possible combinations of data transposes. It can be seen from Fig. \ref{fig:trans} that the $x$ to $z$ transposes and the $z$ to $x$ transposes need a more complex structure, as a process may need to transfer data to other processes which are not immediate neighbours. The non-overlapping of data before and after transposes is most striking for e.g. process $10$ in figure \ref{fig:trans} with no overlap at all between $x$ and $z$ oriented pencils. These transposes are absent in the 2DECOMP library on which we build. These transposes have been implemented using the more flexible all-to-all calls of the type ALLTOALLW, instead of the all-to-all MPI calls of the type ALLTOALLV used for the other four transposes. A complete list of the used libraries can be found in the Appendix.

\begin{table}
\begin{tabular}{r || c | c || c | c ||}
                & \multicolumn{2}{c||}{Slab} & \multicolumn{2}{c||}{Pencil} \\
                \cline{2-5}
                & Halo          & A2A         & Halo           & A2A \\
                \hline
                \hline
$\textbf{u}^*$ computation   & 2             & 6           & 2              & 0  \\
Pressure correction & 8             & 2           & 10             & 6  \\
Scalar equation     & 3             & 2           & 2              & 0  \\
				\hline
Total           & 13            & 10          & 14             & 6  
\end{tabular}
\caption{Number of communications necessary for the computation of all the terms per timestep of the different codes. Here, A2A is short for all-to-all communications. Halo updates involve updating all halo cells, which requires more, but smaller, communications in the case of the pencil code. The difference between the details of the halo and all-to-all for the slab and pencil codes have not been taken into account. It can be seen that the pencil code contains the majority of the communications in the pressure (Poisson) solver. }
\label{tb:comms}
\end{table}

\section{Computational performance}

For optimal scaling conditions, each processor should have an equal amount of work, and the communication to computation ratio should be minimal. In our case, as we do not have iterative solvers, each grid point has the same amount of work, and there is perfect load balancing. We also reduce the communication as far as possible. Not only the number of all-to-all communications are reduced, but also the halo communications as the halo is only one grid point wide. Table \ref{tb:comms} presents the reduction in the number of communications when going from the slab decomposition with fully implicit viscous terms to the pencil decomposition with semi-implicit viscous terms. It is worth noting that the communications are not exactly the same: halo updates involve communications to four neighbours in the pencil decomposition, while only two neighbours are involved in the slab decomposition. However, the size of the halos is relatively smaller, so less data is transferred. Conversely, for the all-to-all communications, not all processes are involved in the pencil code, while all processes take part in the slab code. The memory consumption of the code is approximately $M \approx 15 \times 8 \times N_x \times N_y \times N_z$. Here $M$ is the total used memory in bytes of all processes without MPI overhead. The number of allocated 3D arrays is 14, with additional 1D and 2D arrays of which the memory consumption will not exceed that of one 3D array in the intended cases with large grids. 

\begin{figure}[ht]
    \centering
    \includegraphics[trim=0cm 0cm 0cm 0cm,width=0.47\textwidth]{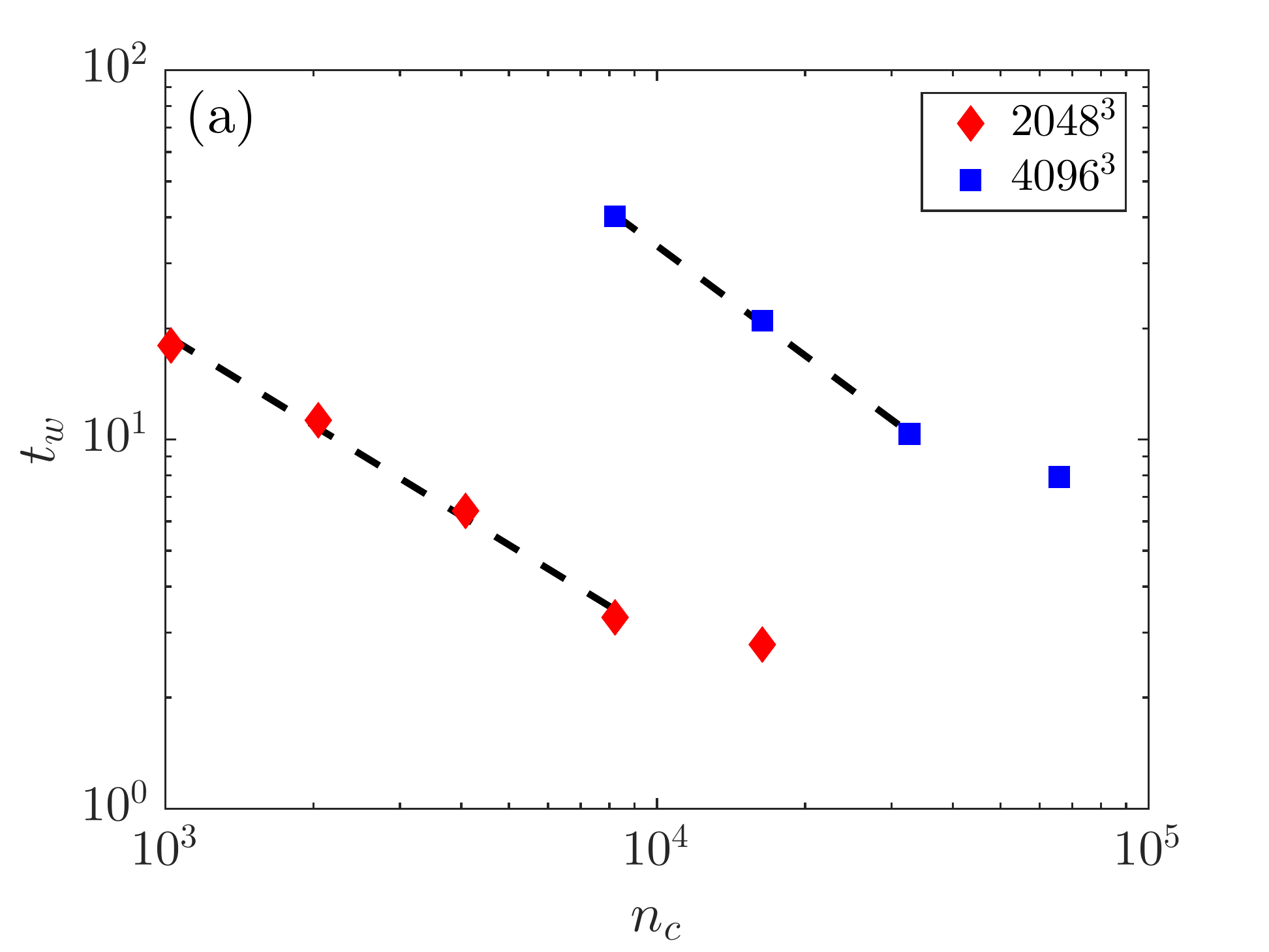}
    \includegraphics[trim=0cm 0cm 0cm 0cm,width=0.47\textwidth]{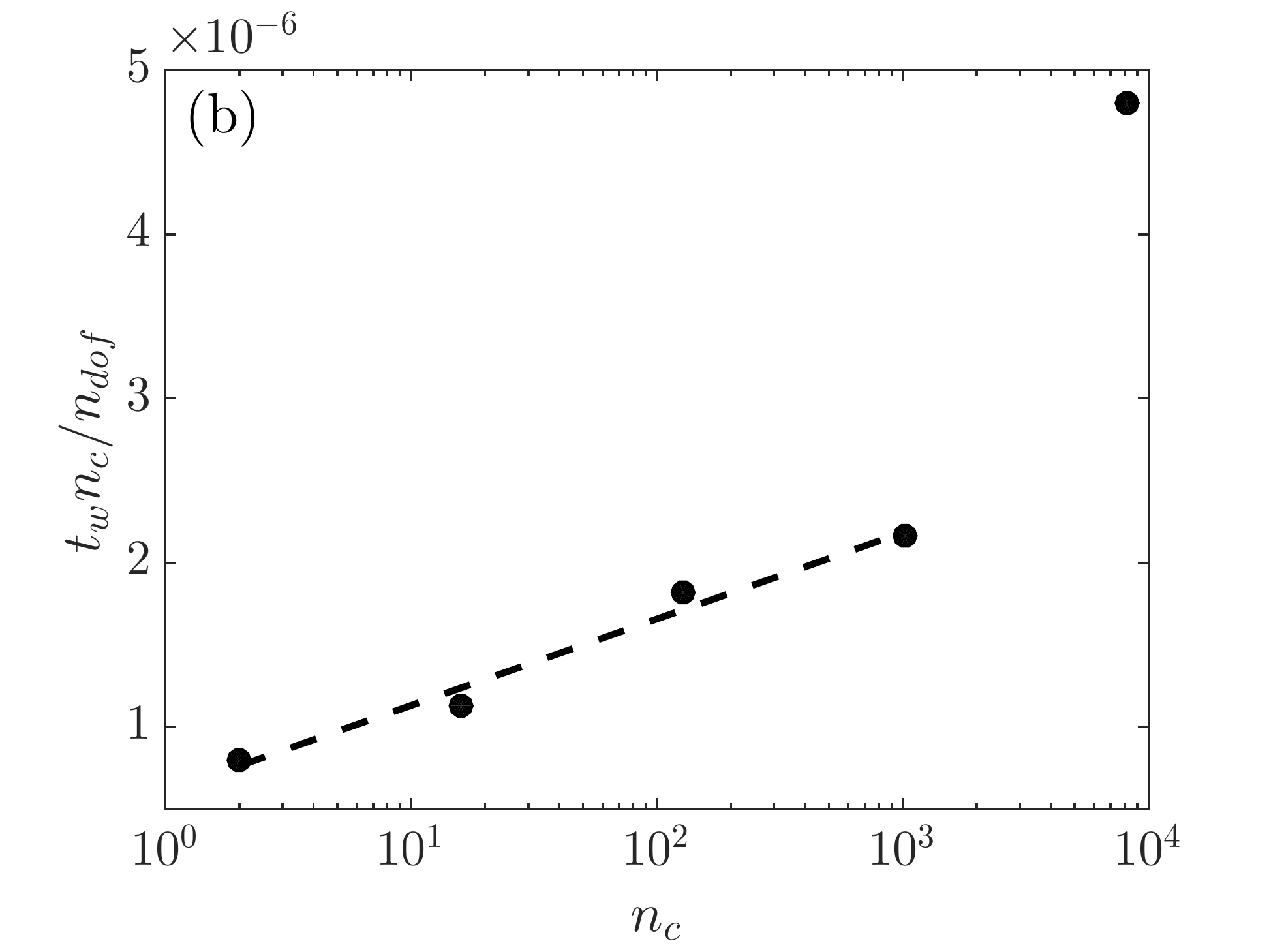}
    \caption{a) Strong scaling of the code for $2048^3$ (red diamonds) and $4096^3$ (blue squares) degrees of freedom $n_{dof}$. Here $n_c$ is the number of cores used. The dashed lines indicate linear behaviour. The walltime per timestep $t_w$ is accounted for a full timestep, i.e. three subtimesteps when using the RK3 integrator. b) Weak scaling of the code for 8.3 Million ($2^{23}$) points per core. The dashed is a linear fit to the corresponding data points.}
\label{fig:scaling}
\end{figure}

The left panel of Figure \ref{fig:scaling} shows the strong scaling of the code for $2048^3$ and $4096^3$ grids on Curie Thin Nodes system. Linear scaling can be seen up to 32K cores for the $4096^3$ grid, with some scaling loss for 64K cores. The right panel of Figure \ref{fig:scaling} shows the weak scaling of the code for $2^{23}$ points from 2 to 16K cores. The data in these plots is obtained by using only MPI parallelism, but hybrid OpenMP/MPI schemes are also available in the code. The choice between pure MPI or hybrid OpenMP/MPI depends heavily on the system on which the code is running. In addition, the presence of OpenMP will allow for a faster porting of the codes to a GPU architecture, in case it becomes viable for our application.

Finally, it must be noted that not only the scaling of the code is excellent, but also the absolute performance. The computational cost per physical time step is very low. In the end, this is what counts.

\section{Summary and prospects}

\begin{figure}
    \centering
    \includegraphics[trim=0cm 0cm 0cm 0cm,width=0.47\textwidth]{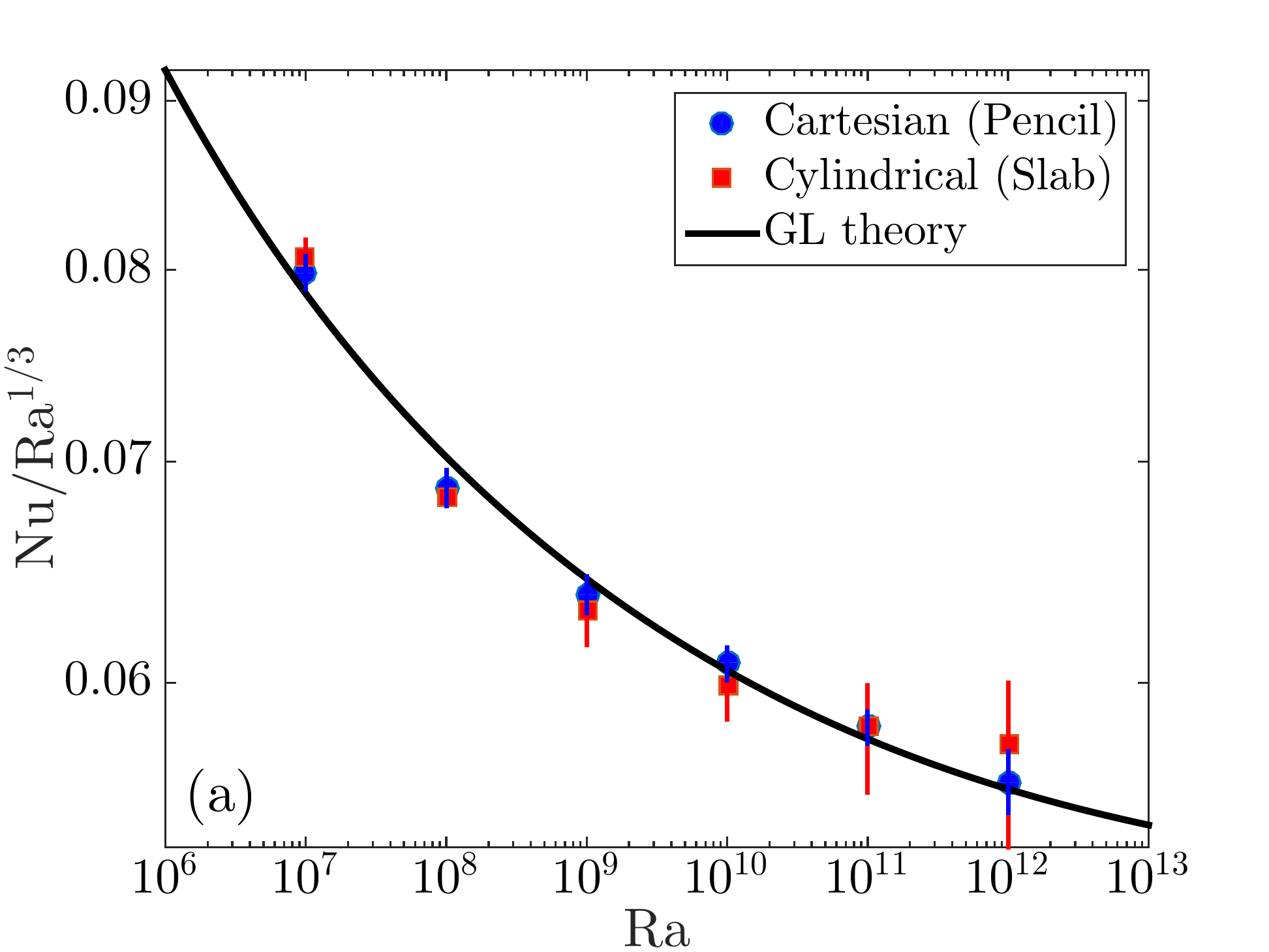}
    \includegraphics[trim=0cm 0cm 0cm 0cm,width=0.47\textwidth]{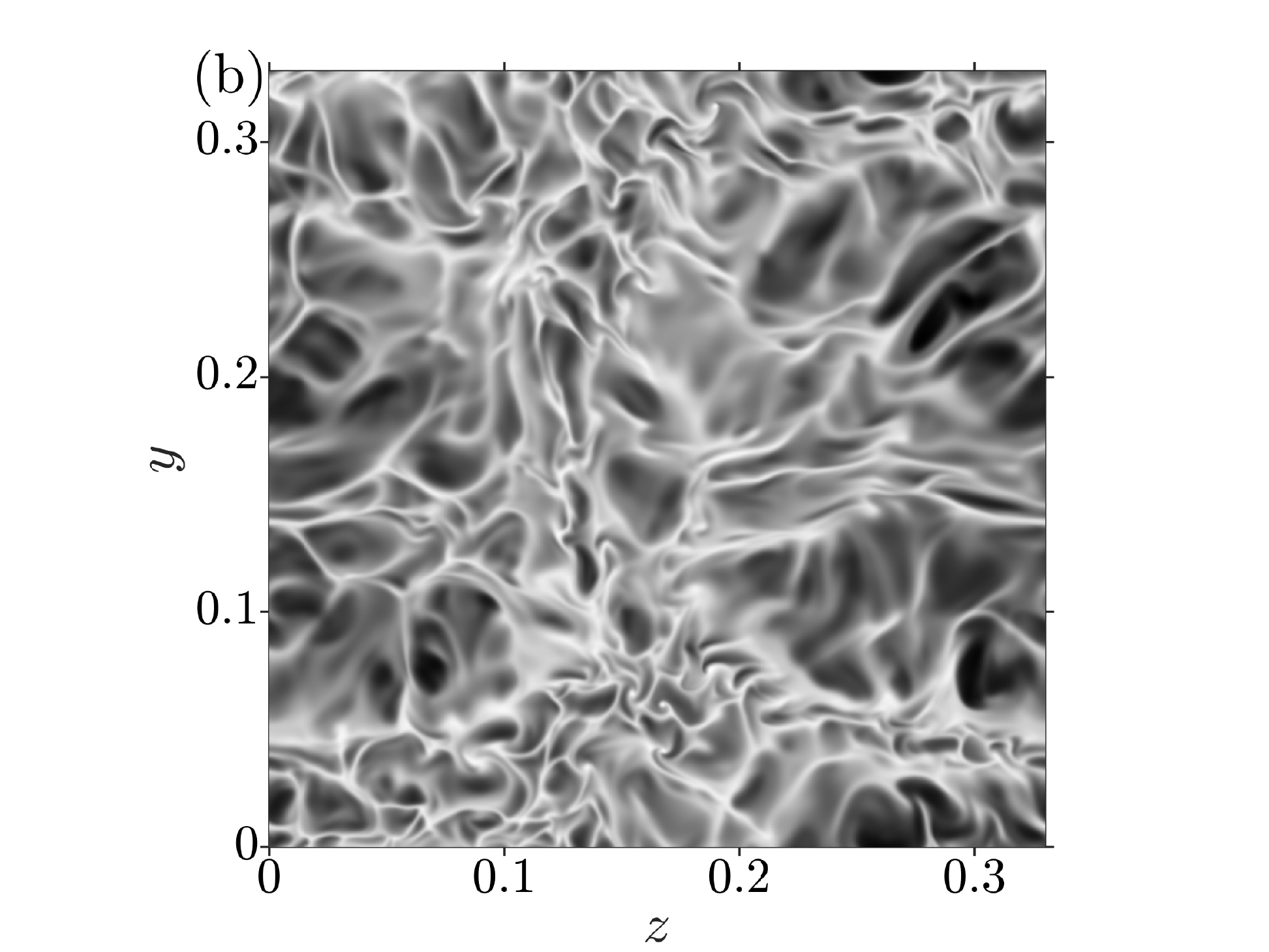}
    \caption{a) The heat flux Nu as a function of the driving Ra for a cylindrical, slab decomposed domain and a Cartesian, pencil decomposed domain. In both cases $\tpr = \mathcal{O}(1)$. The cylinder was simulated with an aspect-ratio of $\Gamma = 1/3$, while the lateral size in the periodic case was sufficiently large to approximate infinite aspect-ratio. The GL theory \cite{gro00} is additionally shown with a solid line. The error bars are based on the temporal convergence of Nu. b) An instantaneous temperature field at height $x = 0.001$ for $\tra = 10^{11}$, where white and black indicate hot and cold fluid, respectively.}
\label{fig:physical}
\end{figure}

In this manuscript, we have presented a parallelization scheme of a second-order centred finite difference method with minimal communication. Only six transposes are needed for every fractional timestep, and for large enough grids, the code's strong scaling is linear up to 32K cores, with slight performance loss from 64K cores.  We emphasize that 64K cores is over half the total number of cores of the Curie Thin nodes system. The code will probably scale well for even larger grids ($8096^3$) on systems with larger amount of cores, as do similar codes based on the 2DECOMP library \cite{li10}. In addition, the absolute performance is excellent. The walltime per physical time is very low and substantial progress can be made with few computational resources. 

The performance of this code allows simulation of flows at high driving. For the application to Rayleigh-B\'enard convection, we refer to figure \ref{fig:physical}a, where the heat flux Nu as a function of the driving Ra can be seen. In this plot, Nu can be compared between the cylindrical, slab decomposed domain, the Cartesian, pencil decomposed domain and the theoretical prediction of the GL-theory \cite{gro00}. The theoretical prediction is based on a fit to experimental data and thus implies an indirect comparison to experiments as well. The cylindrical domain in numerical simulations is used specifically to facilitate a comparison to experiments at the cost of increased time complexity of the pressure correction algorithm and the limitation of a one-dimensional domain decomposition (slab), as compared to the proposed Cartesian code. As a quantification of the difference in computational demands of these code: The highest $\tra = 10^{12}$ data point for the cylindrical and the Cartesian simulations have cost 5M and 1M CPU-hours, respectively. This difference, in favor of the Cartesian method, is amplified by the higher temporal convergence of the Cartesian simulation, judging from the smaller error bar size and the use of more degrees of freedom in the Cartesian geometry, as the system volume is slightly larger for identical $\Gamma$. The heat flux shows negligible differences, which shows that it is largely independent of the sidewall boundary conditions. Even though the impermeable no-slip wall in the cylindrical case is differs largely from the lateral periodicity in the Cartesian case, the heat flux appears unaffected. This indicates that at least for the global quantities, there is no apparent reason to spend more computational resources on a cylindrical simulation, and one can safely resort to the proposed method while maintaining the possibility of comparing to experiments. Even without that possibility, the lateral periodicity is closer to natural applications of RB convection as it approximates infinite aspect-ratio, which by itself warrants the use of the Cartesian domain. The prospected analysis of RB convection is not limited to global quantities such as the heat flux. The highly parallel I/O and high resolution facilitates the study of local quantities. An unfiltered snapshot of the temperature field close to the lower boundary for $\tra = 10^{11}$ is shown in figure \ref{fig:physical}b, where the small scale temperature fluctuations that are required to be properly resolved, can be seen. These small scales can straightforwardly be studied using using spectral analysis, or other techniques. 

The use of this code in Refs. \cite{ost14e,ost15} has already allowed us to push the limits in Taylor-Couette simulations to $\tre\sim\mathcal{O}(10^5)$, never simulated previously. Its use in future RB simulations is expected to allow us to achieve the large driving required for entering the ``ultimate'' regime. The scheme, in combination to a multiple resolution strategy for the scalar field \cite{ost14d}, has been used for simulating double diffusive convection \cite{yan14}, achieving the driving parameters relevant for oceanic convection. Due to the flexibility of finite difference schemes, we expect to be able to build further additions on to this code. The possibility of adding a Lagrangian phase, which can be either tracers, one-way or even two-way coupled particles is detailed in Ref.~\cite{spa14}. Other possibilities include adding rough walls using immersed boundary methods \cite{zhu15}, or adding mixed Neumann-Dirchelet boundary conditions.

\textit{Acknowledgments:} We would like to thank D. Lohse, P. Orlandi, V. Spandan and Y. Yang for useful discussions. We acknowledge FOM, an ERC Advanced Grant and the PRACE grants \#2012061135 and \#2013091966.

\bibliography{/Users/evdp/Documents/PhD/Latex/literatur}

\appendix
\begin{table}
\begin{tabular}{ c | c }
Purpose & Library \\ 
\hline
\hline
I/O & HDF5 \\
\hline
FFT & FFTW (Guru interface) \\
\hline
Linear algebra & BLAS, LAPACK/MKL/ESSL/LibSci/ACML \\
\hline
Distributed memory parallelism & MPI + 2DECOMP \\
\hline 
Shared memory parallelism & OpenMP
\end{tabular}
\caption{The used libraries for the specified purposes are indicated in this table.}
\label{tb:libs}
\end{table}

\end{document}